\documentclass[journal]{IEEEtran}
\usepackage{mathrsfs}
\usepackage{amsmath}
\usepackage{amssymb}
\usepackage{amsthm}
\usepackage{bm}
\ifCLASSINFOpdf \else \fi

\usepackage[dvips]{graphicx}
\usepackage{algorithm}
\usepackage{algorithmic}
\usepackage{slashbox}
\usepackage{subfigure}

\usepackage{color}
\usepackage{slashbox}
\usepackage{multirow}

\usepackage[normalem]{ulem}

\begin{document}

\title{DoA-LF: A Location Fingerprint Positioning Algorithm with Millimeter-Wave}
\author{Zhiqing Wei, Yadong Zhao, Xinyi Liu, Zhiyong Feng
\thanks{The final version of this paper is published in IEEE Access.
Zhiqing Wei, Yadong Zhao, Xinyi Liu
and Zhiyong Feng are with
Beijing University of Posts and Telecommunications,
email: \{weizhiqing, zhaoyadong, xinyi, fengzy\}@bupt.edu.cn.}}


\maketitle

\begin{abstract}
Location fingerprint (LF)
has been widely applied in indoor positioning.
However, the existing studies on LF
mostly focus on the fingerprint of WiFi below 6 GHz,
bluetooth, ultra wideband (UWB), etc.
The LF with millimeter-wave (mmWave) was rarely addressed.
Since mmWave has the characteristics of
narrow beam, fast signal attenuation and wide bandwidth, etc.,
the positioning error can be reduced.
In this paper, an LF
positioning method with mmWave is proposed,
which is named as DoA-LF.
Besides received
signal strength indicator (RSSI) of access points (APs),
the fingerprint database contains
direction of arrival (DoA) information of APs,
which is obtained via DoA estimation.
Then the impact of the number of
APs, the interval of reference points (RPs), the channel model of
mmWave and the error of DoA estimation algorithm on
positioning error is analyzed with
Cramer-Rao lower bound (CRLB).
Finally, the proposed DoA-LF algorithm with mmWave is verified through simulations.
The simulation results have proved that mmWave can reduce
the positioning error
due to the fact that mmWave
has larger path
loss exponent and smaller variance of shadow
fading compared with
low frequency signals.
Besides, accurate DoA estimation can reduce
the positioning error.
\end{abstract}


\IEEEpeerreviewmaketitle
\section{Introduction}


With the development of mobile Internet,
the location based services and applications
such as mobile social networks,
online maps and online-to-offline (O2O) services are
developing rapidly,
which makes the positioning technologies
critical for modern life.
The global navigation satellite system (GNSS) consisting of
GPS, GLONASS, GALILEO, BDS, etc. is efficient in
outdoor environment.
However, in indoor environment, because
satellite signal is blocked by roofs and walls
of buildings, GNSS does not work,
which triggers the studies on indoor
positioning.

\begin{figure}[!t]
\centering
\includegraphics[width=0.48\textwidth]{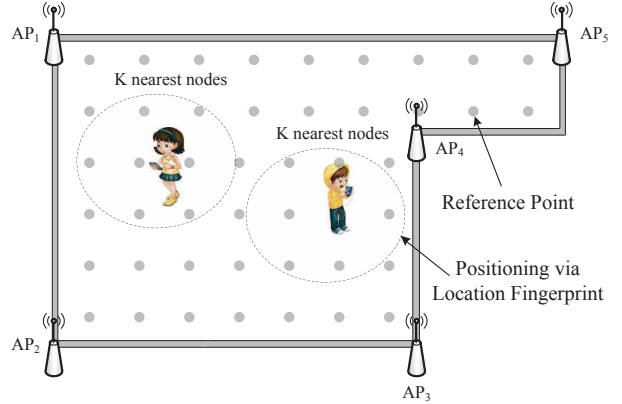}
\caption{The scenario of LF positioning}
\label{fig_scenario}
\end{figure}

LF positioning
plays an important role in complex indoor environment because it
does not require line-of-sight (LOS)
measurement \cite{Survey0}.
The scenario of LF positioning is
illustrated in Fig. \ref{fig_scenario}.
There are five APs
in a specific indoor environment.
The features of received signals
from all APs are measures at
reference points (RPs), which constitute LF database.
When a device appears at a spot,
it measures the features of all APs to
find the RPs with similar features.
Then the estimated location can be
the mean of the locations of $K$
RPs with most similar features,
which is $K$ nearest node (KNN) method.

Among LF positioning systems,
WiFi fingerprint systems are most widely
studied and applied in indoor positioning.
Except of WiFi fingerprint,
UWB fingerprint
systems are also studied \cite{UWB1}\cite{UWB2}.
Besides, with the development of bluetooth low energy (BLE) subsystem,
which is ``designed for machine type communication'' \cite{bluetooth1},
BLE devices will be more and more dense
in buildings. Hence
bluetooth LF
based positioning is also applied in
indoor positioning \cite{bluetooth2}.

The schemes of LF based indoor positioning
focus on
the construction of radio map \cite{radiomap1}\cite{radiomap2},
the contour-based LF method \cite{contour},
the gradient-based fingerprinting \cite{Gradient},
the CSI-based indoor positioning \cite{CSI1}\cite{CSI2}\cite{CSI3} and
machine learning based LF schemes \cite{Kernel}\cite{ELM}, etc.
As to the construction of radio map in LF positioning,
Jiang \emph{et al.} in \cite{radiomap1}
constructed radio map via crowdsourcing
collection.
Jung \emph{et al.} in \cite{radiomap2}
evaluated the performance of
radio map construction methods in various environments.
To provide accurate LF based indoor positioning,
He \emph{et al.} in \cite{contour} developed the contour-based trilateration
to combine the advantages of trilateration and fingerprinting \cite{contour}.
Shu \emph{et al.} in \cite{Gradient} proposed the gradient fingerprinting method to
reduce the impact of varying RSSI
on the accuracy of LF positioning.
LF positioning schemes generally adopt
RSSI information, however, the channel state
information (CSI) contains more information,
which can be
exploited to improve
the accuracy of indoor positioning.
Wu \emph{et al.} in \cite{CSI1}
designed the fingerprinting
system exploiting CSI to
construct the propagation model to improve positioning accuracy.
Wang \emph{et al.} in \cite{CSI2} and \cite{CSI3} designed the
CSI-based fingerprinting schemes via deep learning approach.
The positioning problem is
mathematically a regression problem.
Besides, LF positioning methods have
large amount of data in the offline database.
Thus machine learning approaches can be
adopted to solve the positioning problem.
Mahfouz \emph{et al.} in \cite{Kernel} proposed a
kernel-based machine learning
for LF positioning.
Zou \emph{et al.} applied
extreme learning machine
to solve
indoor positioning problem \cite{ELM}.

Besides the design of
positioning schemes,
the analysis of positioning error is also an active area.
In \cite{RSS_Unknown},
Li proposed RSSI based
positioning scheme without
path loss model and the
Cramer-Rao lower bound (CRLB) was derived to characterize the
positioning error.
Naik \emph{et al.} in \cite{Positioning_Accuracy}
studied the impact of the height of AP
on the positioning accuracy and
the CRLB is correspondingly derived.
Jin \emph{et al.} in \cite{error1}
have studied the performance of RSSI based LF positioning scheme
using CRLB.
And Hossain \emph{et al.} in \cite{error2} analyzed the
CRLB of signal strength difference based LF positioning scheme.

Overall, LF positioning schemes
are mostly focused on
the fingerprint of WiFi, Bluetooth, UWB, etc.
To our best knowledge,
there are rarely studies on
LF positioning with
mmWave.
Because mmWave has the characteristics of
narrow beam, fast signal attenuation and wide bandwidth, etc.,
mmWave signals can provide
centimeter level ranging
accuracy \cite{mmWave0}\cite{mmWave1}\cite{mmWave2}.
Besides, with the development of fifth
generation mobile communication (5G),
dense mmWave small cells will be
widely deployed \cite{5G},
which makes LF
positioning with mmWave to be realizable.
In the standard of IEEE 802.11ad,
60 GHz mmWave for multi-gigabit-per-second
WiFi is developed \cite{IEEE80211ad}.
Besides, IEEE has also started
the standardization of IEEE 802.11aj,
which is a
WLAN system operating at 45 GHz mmWave
band \cite{IEEE80211aj}.
Thus in the near future,
LF positioning with mmWave is practical.

In this paper, an LF positioning method
with mmWave is proposed,
which may be promising in the era of 5G
when APs with mmWave are widely deployed.
We utilize the characteristics of
narrow beam and fast signal attenuation
of mmWave to reduce positioning error.
The proposed LF positioning scheme is called DoA-LF,
because the fingerprint database contains
DoA information
of APs besides RSSI information,
which is obtained via DoA estimation.
Then we select
$K$ candidate RPs with the most similar features
and calculate their weighted mean through
weighted $K$ nearest neighbor (WKNN) algorithm,
which is the final estimated location.
It is noted that the
value of $K$ has an impact on the positioning error and there exists
an optimal $K$ that can minimize the positioning error.
With DoA information of APs operating on mmWave spectrum band,
the positioning error is
significantly reduced compared with LF
positioning with signal below 6 GHz.
Then the impact of the number of APs,
the interval of RPs,
the variance of DoA estimation,
the path loss exponent and
the shadow fading of
mmWave on the positioning error is analyzed via
Cramer-Rao Lower Bound (CRLB).
Finally, the analysis results are
verified by simulation results.
The key parameters and notations in this paper
are listed in Table \ref{sys_para}.

The rest of this paper is organized as follows.
In Section II, system model is introduced.
Section III presents the proposed DoA-LF algorithm.
In Section IV, CRLB is achieved, which
yields the impact of various
parameters on the positioning error.
Simulation results are provided in
Section V to verify our scheme
and analysis results.
Finally, Section VI summarizes
this paper.

\begin{table}[!t]
 \caption{\label{sys_para}Key Parameters and Notations}
 \begin{center}
 \begin{tabular}{l l}
 \hline
 \hline

    {Symbol} & {Description} \\

  \hline

  ${PL({d})}$ & Path loss at a distance $d$\\
  $n$ & Path loss exponent\\
  ${X_\sigma}$ & Shadow fading factor\\
  $(\hat x,\hat y)$ & Estimated location coordinates\\
  $({x_i},{y_i})$ & Location coordinates of $i$th RP\\
  ${{\bm{s}}}$ & Signal strength vector\\
  ${{\bm{\varphi }}}$ & Angle vector\\
  $Q$ & Number of APs\\
  $M$ & Number of RPs\\
  $A$ & Test area\\
  $\theta$ & Location of the test point\\
  $r$ & Location of the estimated point\\
  $\Pr(r|\theta )$ & Conditional probability\\
  ${d_{r\theta }}$ & Euclidean distance between $r$ and $\theta$\\
  ${d_{ir }}$ & Euclidean distance between $r$ and $i$th AP\\
  ${d_{i\theta }}$ & Euclidean distance between $\theta$ and $i$th AP\\
  $\bm{s}_{i\theta}$ & Actual signal strength of $i$th AP at the test point\\
  $\bm{s}_{ir}$ & Received signal strength of the $i$th AP at point $r$\\
  ${\varphi _{i\theta}}$ & DoA of $i$th AP at the test point\\
  ${\varphi _{ir}}$ & DoA of the $i$th AP at point $r$\\
  ${\sigma _{s}^2}$ & Variance of received signal strength\\
  ${\sigma _{\varphi}^2}$ & Variance of DoA estimation\\
  $J(\hat \theta)$ &  Fisher Information Matrix\\
  ${C_{\hat \theta }}$ & Cramer-Rao Lower Bound\\
  AP & Access point\\
  RP & Reference point\\
  TP & Test point\\
  LF & Location fingerpint\\
  RSSI & Received signal strength indicator\\
  \hline
  \hline
 \end{tabular}
 \end{center}
\end{table}

\section{System Model}

The 60 GHz mmWave is adopted for
multi-gigabit-per-second WiFi \cite{IEEE80211ad}.
Hence 60 GHz mmWave is adopted in LF positioning.
In this section, the channel model of 60 GHz
mmWave and the process of
LF positioning are presented.

\subsection{60 GHz mmWave Channel Model}

The channel model of
60 GHz mmWave is \cite{9}\cite{9r}

\begin{equation}
PL(d)[dB] = PL({d_0}) + 10n\log (\frac{d}{{{d_0}}}) + {X_\sigma },d > {d_0},
\label{e1}
\end{equation}
where $PL({d_0}) = 20\log (\frac{{4\pi {d_0}}}{\lambda })$ is
free-space path loss
at a reference distance ${d_0}$,
which is generally set as ${d_0} = 1 \kern 2pt\rm{m}$.
The path loss exponent is $n$.
${X_\sigma}$ is the shadow fading factor
modeled by a Gaussian random variable
with mean zero and variance $\sigma_s^2$.

In DOA-LF algorithm, besides RSSI information,
DoA information is also needed, which can be obtained via
multiple signal classification (MUSIC)
algorithm.
Since MUSIC algorithm is well known
in the area of array signal
processing, we do not introduce
MUSIC algorithm in this paper.
Readers can refer to \cite{MUSIC} for details.
The MUSIC algorithm has
lower computational complexity compared with other DoA algorithms.
However, other DoA algorithms can also
be adopted besides MUSIC algorithm.
If a DoA algorithm has small error,
the performance of positioning algorithm can
be correspondingly improved.

\subsection{DoA-LF Positioning Algorithm}
LF positioning
plays an important role
in complex indoor environment.
LF positioning algorithm consists of two steps, namely,
offline database construction and online matching.
Weighted $K$ nearest neighbor
(WKNN) algorithm is widely
applied in LF positioning \cite{WKNN0}\cite{WKNN1}\cite{WKNN2},
where the weighted
mean of the coordinates of
$K$ nearest RPs is calculated
as the estimated coordinates.
The weight coefficient $w_i$ is
generally inversely proportional to
the Euclidean distance between the
estimated point and $i$th RP \cite{WKNN0}\cite{WKNN1}
in the feature space spanning
by the vectors containing
RSSI and DoA information.
Hence the estimated
coordinates are as follows \cite{WKNN0}\cite{WKNN1}.

\begin{equation}\label{eq_estimation}
\begin{aligned}
& {w_i} = \frac{\gamma }{{{d_i} + \varepsilon }},\\
& (\hat x,\hat y) = \frac{{\sum\limits_{i = 1}^K {{w_i}({x_i},{y_i})} }}{{\sum\limits_{i = 1}^K {{w_i}} }},
\end{aligned}
\end{equation}
where ${d_i}$ is the Euclidean distance in the feature space between
the measured point and the $i$th RP.
$(\hat x,\hat y)$ are the estimated coordinates
and $({x_i},{y_i})$ are the coordinates of $i$th RP.
$\gamma $ is normalized parameter
and $\varepsilon $ is a small positive number
in order to prevent the denominator being zero.

\section{The Proposed DoA-LF Algorithm}

\begin{figure}[!t]
\centering
\includegraphics[width=0.45\textwidth]{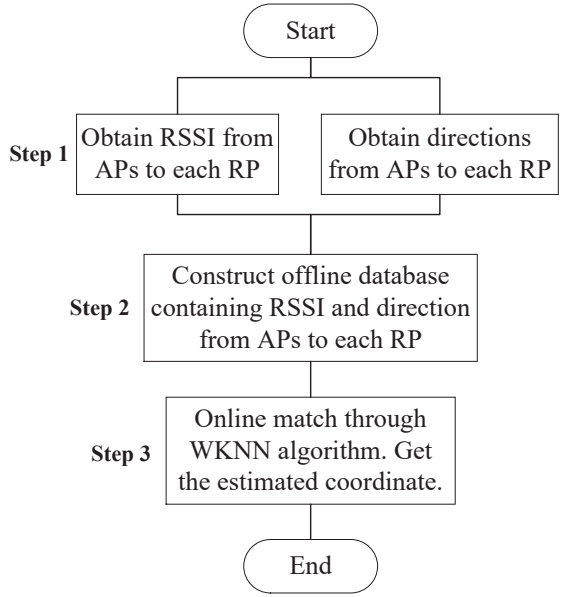}
\caption{The proposed DoA-LF algorithm.}
\label{fig_frame_structure1}
\end{figure}

In this section, the characteristics
of mmWave, such as narrow beam and fast attenuation, are
exploited to construct an LF positioning algorithm
called DoA-LF algorithm.
The DoA information of APs
obtained by MUSIC algorithm
is combined with the RSSI of APs to construct a
new offline database for online matching.
Then WKNN algorithm is adopted to calculate
the estimated coordinates.
Fig. \ref{fig_frame_structure1} illustrates the process of proposed DoA-LF algorithm.

\textbf{Step 1}: Obtain RSSI and DoA information from APs to each RP.

\begin{figure}[!t]
\centering
\includegraphics[width=0.45\textwidth]{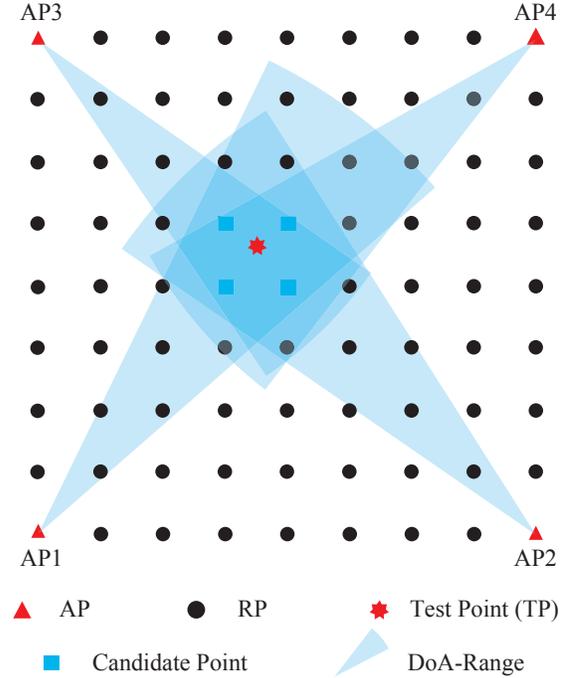}
\caption{A scenario of proposed DoA-LF positioning.}
\label{fig_scenarsio}
\end{figure}

$Q$ APs and $M$ RPs are deployed in a specific area,
which is illustrated in Fig. \ref{fig_scenarsio}.
The RSSIs of APs at each RP are measured,
which are saved in a
vector ${{\bm{s}_i}} = [{s_{1i}},{s_{2i}}, \cdots ,{s_{Qi}}]$ with
${s_{ji}},j = 1, \cdots ,Q$ denoting the RSSI of $j$th AP at $i$th RP.

Meanwhile,
the MUSIC algorithm is adopted for DoA estimation.
The directions of APs at each RP are measured,
which are saved in a vector
${{\bm{\varphi}_i}} = [{\varphi _{1i}},{\varphi _{2i}}, \cdots ,{\varphi _{Qi}}]$ with
${\varphi _{ji}},j = 1, \cdots ,Q$
denoting the DoA of $j$th AP at $i$th RP.
It is noted that the RSSI and DoA information
is obtained simultaneously.

\textbf{Step 2}: Construct offline database.

Combining $Q$-dimensional RSSI
vector and $Q$-dimensional
direction vector,
we obtain $2Q$-dimensional
vector $[\bm{s}_i, \bm{\varphi}_i]$
representing the features of
$i$th RP.
The vectors of all RPs span a feature space and
construct the offline
database for DoA-LF positioning with mmWave.

\textbf{Step 3}: Online matching.

Comparing RSSI and direction
information of test point
with the data in offline database,
we can find out
$K$ nearest RPs
whose features are closest to test point.
Then we calculate the estimated coordinates via
the weighted mean
of $K$ nearest RPs using (\ref{eq_estimation}).

For example, in Fig. \ref{fig_scenarsio},
there are $4$ APs and $77$ RPs
in an area.
Firstly,
an offline database is established,
which consists of
RSSI and DoA information of all RPs.
Secondly, $K = 4$ candidate RPs with
the most similar features are selected,
which
are denoted by the
squares
in Fig. \ref{fig_scenarsio}.
Finally,
the estimated coordinates can be obtained
via calculating the weighted mean
of the selected $K = 4$
candidate RPs.

\section{Analysis of Positioning Error}
\label{sec_CRLB}

In the above section,
the DoA-LF positioning algorithm with mmWave is proposed.
In DoA-LF positioning algorithm,
the number of APs, the interval of RPs,
the channel model of mmWave and the error of
DoA estimation have an
impact on the positioning error.
In this section,
we analyze the impact of these parameters
on positioning error,
which can provide guideline
for the selection of
appropriate parameters
in DoA-LF algorithm with mmWave.

As illustrated in Fig. \ref{fig_scenarsio},
there are $Q$
APs and $M$ RPs.
APs are distributed on the edge of a specific area
denoted as area $A$.
RPs are uniform distributed
in area $A$.
The location of test point
is denoted as $\theta$.
DoA-LF algorithm generates
the estimated point $r$ with conditional
probability $\Pr(r|\theta )$.
The Euclidean
distance between $r$ and $\theta$ is ${d_{r\theta }}$.
The conditional probability $\Pr(r|\theta )$
and the distance ${d_{r\theta }}$ can
represent positioning error. Besides,
$\Pr(r|\theta )$ is related with ${d_{r\theta }}$.
In LF positioning, one-to-one mapping can be established
between $2Q$-dimensional feature space and two-dimensional
geo-location space.
Thus the expression of conditional
probability $\Pr(r|\theta )$ in two-dimensional
geo-location space equals to the conditional probability
$\Pr({\bm{s}_{r}},{\bm{\varphi}_{r}}|{\bm{s}_{\theta }},{\bm{\varphi}_{\theta }})$
in $2Q$-dimensional feature space,
where ${\bm{s}_{\theta }}$, ${\bm{\varphi}_{\theta }}$,
${\bm{s}_{r}}$ and ${\bm{\varphi}_{r}}$ are the
RSSI and DoA information of all APs
at location $\theta$ and $r$ respectively.
Therefore the conditional
probability $\Pr(r|\theta )$
is as follows.

\begin{equation}
\begin{aligned}
\Pr(r|\theta ) & = \Pr({\bm{s}_{r}},{\bm{\varphi}_{r}}|{\bm{s}_{\theta }},{\bm{\varphi}_{\theta }})\\
& = \prod\limits_{i = 1}^Q {\Pr({\bm{s}_{ir}},{\bm{\varphi}_{ir}}|{\bm{s}_{i\theta }},{\bm{\varphi}_{i\theta }})} \\
& = \prod\limits_{i = 1}^Q {\Pr({\bm{s}_{ir}}|{\bm{s}_{i\theta }})} \prod\limits_{i = 1}^Q {\Pr({\bm{\varphi}_{ir}}|{\bm{\varphi}_{i\theta }})}.
\end{aligned}
\label{e11}
\end{equation}

The conditional probability $\Pr({\bm{s}_{ir}}|{\bm{s}_{i\theta }})$
denotes that the actual signal
strength of $i$th AP at the test point is $\bm{s}_{i\theta}$
while the received signal strength of $i$th AP is $\bm{s}_{ir}$,
which is the signal strength at location $r$.
$\Pr({\bm{s}_{ir}}|{\bm{s}_{i\theta }})$
follows Gaussian distribution with mean $\bm{s}_{i\theta}$
and variance ${\sigma _{s}^2}$ \cite{error2}\cite{16}.
Meanwhile, $\Pr({\bm{\varphi}_{ir}}|{\bm{\varphi}_{i\theta }})$
denotes the conditional probability that the
DoA of $i$th AP at test point is ${\varphi _{i\theta}}$
while the estimated DoA of $i$th AP
is ${\varphi _{ir}}$, which is the DoA
of $i$th AP at location $r$.
$\Pr({\bm{\varphi}_{ir}}|{\bm{\varphi}_{i\theta }})$ also follows
Gaussian distribution with mean $\bm{\varphi}_{i\theta}$
and variance ${\sigma _{\varphi}^2}$ \cite{17}\cite{18}.

\begin{figure*}[!t]
\normalsize
\begin{equation}  \label{e122}
\begin{aligned}
&{ \frac{{{\partial ^2}\log f(r;\theta )}}{{\partial x^2}}{\rm{ = }}\sum\limits_{i = 1}^Q {\left( {\frac{{ - 2{\eta}{{\cos }^2}{\varphi _{i\theta }}}}{{d_{i\theta }^2}} + {\eta}\log (\frac{{d_{ir}^2}}{{d_{i\theta }^2}})\frac{{1 - 2{{\cos }^2}{\varphi _{i\theta }}}}{{d_{i\theta }^2}} - \frac{{si{n^2}{\varphi _{i\theta }}}}{{\sigma _\varphi ^2d_{i\theta }^2}}} \right)}}, \\
&{\frac{{{\partial ^2}\log f(r;\theta )}}{{\partial {x}\partial {y}}}{\rm{ = }}\frac{{{\partial ^2}\log f(r,\theta )}}{{\partial {y}\partial {x}}}{\rm{ = }}\sum\limits_{i = 1}^Q {\left( { - \frac{{{\eta}sin2{\varphi _{i\theta }}}}{{d_{i\theta }^2}} - {\eta}\log (\frac{{d_{ir}^2}}{{d_{i\theta }^2}})\frac{{{\rm{sin}}2{\varphi _{i\theta }}}}{{d_{i\theta }^2}} + \frac{{({\varphi _{ir}} - {\varphi _{i\theta }})cos2{\varphi _{i\theta }} - cos{\varphi _{i\theta }}\sin {\varphi _{i\theta }}}}{{\sigma _\varphi ^2d_{i\theta }^2}}} \right)}}, \\
&{\frac{{{\partial ^2}\log f(r,\theta )}}{{\partial y^2}}{\rm{ = }}\sum\limits_{i = 1}^Q {\left( {\frac{{ - 2{\eta}si{n^2}{\varphi _{i\theta }}}}{{d_{i\theta }^2}} + {\eta}\log (\frac{{d_{ir}^2}}{{d_{i\theta }^2}})\frac{{1 - 2{{\sin }^2}{\varphi _{i\theta }}}}{{d_{i\theta }^2}} - \frac{{co{s^2}{\varphi _{i\theta }}}}{{\sigma _\varphi ^2d_{i\theta }^2}}} \right)}}.
\end{aligned}
\end{equation}
\hrulefill \vspace*{4pt}
\end{figure*}

According to the path loss
model of mmWave in (\ref{e1}),
assuming that the transmit power of each
AP is $P_t$, the average signal
strength at $\theta $ from $i$th AP is given by
\begin{equation}
{s_{i\theta }} = P_t - PL({d_{i\theta }}) = P_t - PL({d_0}) - 10n\log (\frac{{{d_{i\theta }}}}{{{d_0}}}) - {X_\sigma},
\label{e13}
\end{equation}
where $X_\sigma$ is a Gaussian random variable
with mean zero and variance $\sigma_s^2$.
$d_{i\theta}$ is the distance between $i$th AP and the location $\theta$.

CRLB is an
effective tool to estimate
the minimum variance of parameter
estimation error \cite{19}\cite{20}\cite{21}.
In this subsection,
the lower
bound of the variance of DoA-LF algorithm error is analyzed
by CRLB.
Assuming that the test
point is $\theta = (x,y)^T$,
the unbiased estimation value
of $\theta$ is $\hat \theta  = (\hat x,\hat y)^T$,
which is
the weighted mean of $K$ nearest RPs
selected by DoA-LF algorithm.
The covariance matrix of $\hat \theta$ is

\begin{equation}
\begin{aligned}
& {\mathop{\rm cov}} \left( \hat \theta \right) = E\left( {(\hat \theta  - \theta ){{(\hat \theta  - \theta )}^T}} \right)\\
& = \left( {\begin{array}{*{20}{c}}
{{\mathop{\rm var}} (\hat x - x)}&{{\mathop{\rm cov}} \left( {(\hat x - x),(\hat y - y)} \right)}\\
{{\mathop{\rm cov}} \left( {(\hat y - y),(\hat x - x)} \right)}&{{\mathop{\rm var}} (\hat y - y)}
\end{array}} \right),
\end{aligned}
\label{e15}
\end{equation}
where $E(\alpha)$
is the expectation of random variable $\alpha$.
${\mathop{\rm var}} (\alpha )$ is the variance of $\alpha$.
$\rm{cov}(\alpha ,\beta )$ is the covariance
of random variables $\alpha$ and $\beta$.
According to CRLB inequality,
the covariance of $\hat \theta$
satisfies the following inequality \cite{book_estimation}\cite{23}.

\begin{equation}
{\mathop{\rm cov}} (\hat \theta) \ge \left( J(\hat \theta)\right) ^{ - 1},
\label{e16}
\end{equation}
where $J(\hat \theta)$ is fisher information matrix (FIM),
which is defined as follows \cite{20}\cite{23}.

\begin{equation}
J(\hat \theta) = {E}\left(  - \frac{{{\partial ^2}\ln f(r;\theta )}}{{\partial \theta \partial {\theta ^T}}}\right),
\label{e17}
\end{equation}
where $r$ is
the corresponding coordinates of
$2Q$-dimensional observation features and
$f(r;\theta )$ is denoted as

\begin{equation}
f(r;\theta ) = f(r|\theta )f(\theta ).
\label{e118}
\end{equation}

According to one-to-one mapping relation between
$2Q$-dimensional feature space and
two-dimensional geo-location space, the value of
$f(r|\theta )$ is provided in (\ref{e11}).
Besides, $\theta$ is
uniformly distributed in area $A$,
hence $f(\theta ) = \frac{1}{{\left| A \right|}}$
where $\left| A \right|$ denotes the area of $A$.
Thus the value of $f(r;\theta )$ is

\begin{equation}
\begin{aligned}
& f(r;\theta ) = \prod\limits_{i = 1}^Q {{\kappa}\exp \left( { - \frac{{{{({s_{ir}} - {s_{i\theta }})}^2}}}{{2\sigma _{s}^2}} - \frac{{{{({\varphi _{ir}} - {\varphi _{i\theta }})}^2}}}{{2\sigma _{\varphi }^2}}} \right)}, \\
& = \prod\limits_{i = 1}^Q {{\kappa}\exp \left( { - \frac{{{{\left( {10n\lg (\frac{{{d_{ir}}}}{{{d_{i\theta }}}})} \right)}^2}}}{{2\sigma _{s}^2}} - \frac{{{{({\varphi _{ir}} - {\varphi _{i\theta }})}^2}}}{{2\sigma _{\varphi}^2}}} \right)},
\end{aligned}
\label{e119}
\end{equation}
where $\kappa  = \frac{1}{{2\pi |A|{\sigma _{s }}{\sigma _{\varphi }}}}$.
$d_{ir}$ is the distance between $i$th AP and the location $r$.
$\lg (*)$ is a logarithmic function with base $10$.
Therefore the function
$\log f(r;\theta )$  is as follows.

\begin{equation}
\begin{aligned}
& \log f(r;\theta )\\
 &= \sum\limits_{i = 1}^Q {\log \left( {{\kappa}\exp \left( - \frac{{{{(10n\lg (\frac{{{d_{ir}}}}{{{d_{i\theta }}}}))}^2}}}{{2\sigma _s^2}} - \frac{{{{({\varphi _{ir}} - {\varphi _{i\theta }})}^2}}}{{2\sigma _\varphi ^2}}\right)} \right)} \\
& = \sum\limits_{i = 1}^Q {\left( {{\rm{log}}{\kappa} - {\eta}{{\left(\log (\frac{{{d_{ir}}}}{{{d_{i\theta }}}})\right)}^2} - \frac{{{{({\varphi _{ir}} - {\varphi _{i\theta }})}^2}}}{{2\sigma _\varphi ^2}}} \right)},
\end{aligned}
\label{e120}
\end{equation}
where $\log (*)$ is a logarithmic function with base $e$.
The value of ${\eta}$ is

\begin{equation}\label{eq_eta}
{\eta} = {(\frac{{10n}}{{\sqrt 2 {\sigma_s } \log 10}})^2}.
\end{equation}

\begin{figure}[!t]
\centering
\includegraphics[width=0.48\textwidth]{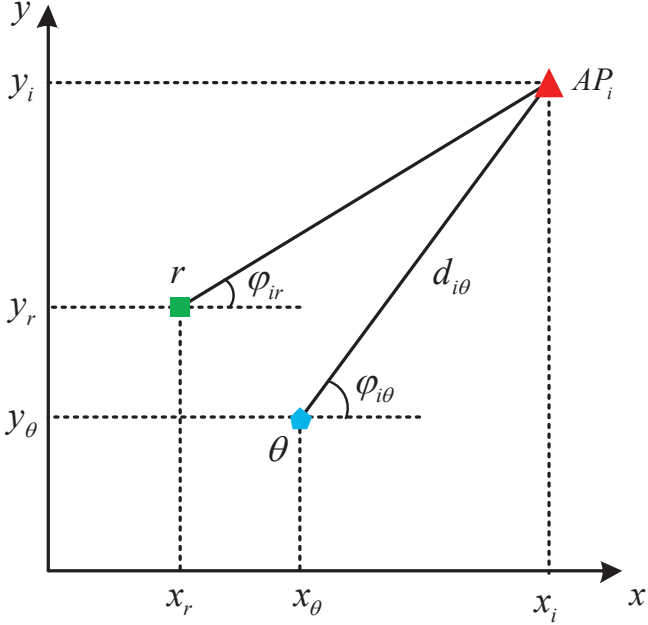}
\caption{The distance and DoA of $\theta$ and $r$.}
\label{f3}
\end{figure}

It is noted that
${d_{i\theta }}$ and ${\varphi _{i\theta }}$ are
functions of
$x$ and $y$,
whose expressions are

\begin{equation}
\begin{aligned}
& {d_{i\theta }}{\rm{ = }}\sqrt {{{({x} - {x_i})}^2} + {{({y} - {y_i})}^2}},\\
& {\varphi _{i\theta }} = \arccos (\frac{{{x_i} - {x}}}{{{d_{i\theta }}}}) = \arcsin (\frac{{{y_i} - {y}}}{{{d_{i\theta }}}}),
\end{aligned}
\end{equation}
which are illustrated by Fig. \ref{f3}.

In order to obtain (\ref{e17}), we first derive
the first order partial derivatives as follows.

\begin{equation}
\begin{aligned}
&\frac{{\partial \log f(r;\theta )}}{{\partial {x}}} = \sum\limits_{i = 1}^Q {\eta \log (\frac{{d_{ir}^2}}{{d_{i\theta }^2}})\frac{{{x} - {x_i}}}{{d_{i\theta }^2}} + \frac{{({\varphi _{ir}} - {\varphi _{i\theta }})\sin {\varphi _{i\theta }}}}{{\sigma _\varphi ^2{d_{i\theta }}}}},\\
&\frac{{\partial \log f(r;\theta )}}{{\partial {y}}} = \sum\limits_{i = 1}^Q {\eta \log (\frac{{d_{ir}^2}}{{d_{i\theta }^2}})\frac{{{y} - {y_i}}}{{d_{i\theta }^2}} + \frac{{({\varphi _{ir}} - {\varphi _{i\theta }})\cos {\varphi _{i\theta }}}}{{\sigma _\varphi ^2{d_{i\theta }}}}}.
\end{aligned}
\label{e121}
\end{equation}

Then the second order derivatives
can be derived,
which are shown in (\ref{e122}).

For DoA-LF algorithm, ${d_{ir}}$
approximately
equals ${d_{i\theta }}$
and ${\varphi _{ir}}$
approximately equals ${\varphi _{i\theta }}$.
Hence the values
of $\log (\frac{{d_{ir}^2}}{{d_{i\theta }^2}})$
and ${\varphi _{ir}} - {\varphi _{i\theta }}$ approximate zero.
The entries of (\ref{e122}) become as follows.

\begin{equation}
\begin{aligned}
& \frac{{{\partial ^2}\log f(r;\theta )}}{{\partial x^2}} = \sum\limits_{i = 1}^Q {\left( { - \frac{{2\eta {{\cos }^2}{\varphi _{i\theta }}}}{{d_{i\theta }^2}} - \frac{{{{\sin }^2}{\varphi _{i\theta }}}}{{\sigma _\varphi ^2d_{i\theta }^2}}} \right)}, \\
& \frac{{{\partial ^2}\log f(r;\theta )}}{{\partial {x}\partial {y}}} =  \frac{{{\partial ^2}\log f(r;\theta )}}{{\partial {y}\partial {x}}}\\
&{\kern 57pt} = \sum\limits_{i = 1}^Q {( - \frac{{{\eta}\sin 2{\varphi _{i\theta }}}}{{d_{i\theta }^2}} - \frac{{\sin 2{\varphi _{i\theta }}}}{{2\sigma _\varphi ^2d_{i\theta }^2}})}, \\
& \frac{{{\partial ^2}\log f(r;\theta )}}{{\partial y^2}} = \sum\limits_{i = 1}^Q {( - \frac{{2{\eta}{{\sin }^2}{\varphi _{i\theta }}}}{{d_{i\theta }^2}} - \frac{{{{\cos }^2}{\varphi _{i\theta }}}}{{\sigma _{\varphi }^2d_{i\theta }^2}})}.
\end{aligned}
\label{e123}
\end{equation}

Then the FIM $J(\hat \theta )$ can be derived as follows.

\begin{equation}
J(\hat \theta ) = \left( {\begin{array}{*{20}{c}}
{{J_{xx}}(\hat \theta )}&{{J_{xy}}(\hat \theta )}\\
{{J_{yx}}(\hat \theta )}&{{J_{yy}}(\hat \theta )}
\end{array}} \right),
\label{e124}
\end{equation}
where the entries of (\ref{e124}) are

\begin{equation}
\begin{aligned}
& {J_{{x}{x}}}(\hat \theta ) =  - \frac{{{\partial ^2}\log f(r;\theta )}}{{\partial x^2}} = \sum\limits_{i = 1}^Q {\frac{{2\eta {{\cos }^2}{\varphi _{i\theta }}}}{{d_{i\theta }^2}} + \frac{{{{\sin }^2}{\varphi _{i\theta }}}}{{\sigma _\varphi ^2d_{i\theta }^2}}},\\
& {J_{xy}}(\hat \theta ) = {J_{yx}}(\hat \theta ) = \sum\limits_{i = 1}^Q {\frac{{{\eta}\sin 2{\varphi _{i\theta }}}}{{d_{i\theta }^2}} + \frac{{\sin 2{\varphi _{i\theta }}}}{{2\sigma _\varphi ^2d_{i\theta }^2}}},\\
& {J_{yy}}(\hat \theta ) =  - \frac{{{\partial ^2}\log f(r;\theta )}}{{\partial y^2}} = \sum\limits_{i = 1}^Q {\frac{{2{\eta}{{\sin }^2}{\varphi _{i\theta }}}}{{d_{i\theta }^2}} + \frac{{{{\cos }^2}{\varphi _{i\theta }}}}{{\sigma _\varphi ^2d_{i\theta }^2}}}.
\end{aligned}
\label{e125}
\end{equation}

Therefore ${\left( {J(\hat \theta )} \right)^{ - 1}}$ is

\begin{equation}
{\left( {J(\hat \theta )} \right)^{ - 1}} = \frac{1}{{\left| {J(\hat \theta )} \right|}}\left( {\begin{array}{*{20}{c}}
{{J_{yy}}(\hat \theta )}&{ - {J_{yx}}(\hat \theta )}\\
{ - {J_{xy}}(\hat \theta )}&{{J_{xx}}(\hat \theta )}
\end{array}} \right),
\label{e126}
\end{equation}
where $\left| {J(\hat \theta )} \right|$ is

\begin{equation}
\begin{aligned}
\left| {J(\hat \theta )} \right| & = {J_{xx}}(\hat \theta ){J_{yy}}(\hat \theta ) - {J_{xy}}(\hat \theta ){J_{yx}}(\hat \theta )\\
& { = \sum\limits_{i = 1}^Q {\frac{{2{\eta}{{({{\cos }^2}{\varphi _{i\theta }} - {{\sin }^2}{\varphi _{i\theta }})}^2}}}{{\sigma _{\varphi }^2d_{i\theta }^4}}} }.
\end{aligned}
\label{e127}
\end{equation}

Substituting (\ref{e15}) and (\ref{e126}) into
(\ref{e16}), we have

\begin{equation}
\begin{aligned}
& \left( {\begin{array}{*{20}{c}}
{{\mathop{\rm var}} (\hat x - x)}&{{\mathop{\rm cov}} \left( {(\hat x - x),(\hat y - y)} \right)}\\
{{\mathop{\rm cov}} \left( {(\hat y - y),(\hat x - x)} \right)}&{{\mathop{\rm var}} (\hat y - y)}
\end{array}} \right)\\
& \ge \frac{1}{{|J(\hat \theta )|}}\left( {\begin{array}{*{20}{c}}
{{J_{yy}}(\hat \theta )}&{ - {J_{yx}}(\hat \theta )}\\
{ - {J_{xy}}(\hat \theta )}&{{J_{xx}}(\hat \theta )}
\end{array}} \right).
\end{aligned}
\label{e128}
\end{equation}

Hence the CRLB ${C_{\hat \theta }}$
of positioning error is

\begin{equation}
\begin{aligned}
{C_{\hat \theta }} & = {{\mathop{\rm var}} (\hat x - x)}  + {{\mathop{\rm var}} (\hat y - y)} \ge \frac{{{J_{xx}}(\hat \theta ) + {J_{yy}}(\hat \theta )}}{{|J(\hat \theta )|}}\\
& = \sum\limits_{i = 1}^Q {\frac{{d_{i\theta }^2(1 + 2{\eta}\sigma _\varphi ^2)}}{{2{\eta}{{({{\cos }^2}{\varphi _{i\theta }} - {{\sin }^2}{\varphi _{i\theta }})}^2}}}}.
\end{aligned}
\label{e129}
\end{equation}

According to (\ref{e129}),
CRLB is an increasing function of $\sigma _{\varphi}^2$.
Besides, CRLB is a decreasing function of ${\eta}$.
Because of (\ref{eq_eta}), we have conclusions as follows.

\begin{enumerate}
  \item The variance of positioning error is increasing with
the variance $\sigma_s^2$, which means
that the positioning error is increasing with the increase of
shadow fading factor.
  \item The variance of positioning error is decreasing with
path loss exponent $n$, which
means that LF positioning error can be
reduced when using
the signal with large
path loss exponent, such as mmWave.
 \item In LF positioning with mmWave, when $\sigma _{\varphi}^2$
 is small, namely, the error of DoA estimation is small, the
 positioning error can be reduced.
\end{enumerate}

Meanwhile, the variance of
positioning error is also impacted by
the interval of RPs
and the number of APs.
With the decreasing of
the interval of RPs,
the variance of positioning error is decreasing.
However, when the interval is extremely small,
the computational complexity of
DoA-LF algorithm will be large.
Besides, with the increasing of
the number of APs,
the positioning error can be reduced.
However, when the number of APs is extremely large,
the computational complexity of DoA-LF algorithm will
be intolerable.
Simulation results in
Section \ref{sec_simulation} will
support our analysis.

\section{Simulation Results and Analysis}
\label{sec_simulation}

\subsection{DoA-LF Algorithm}

Firstly, the LF positioning scheme
with mmWave is compared with the
traditional LF positioning with 2.4 GHz WiFi.
The path loss models of 60 GHz mmWave and 2.4 GHz WiFi
under the same
indoor NLOS\footnote{The LF positioning is mainly applied in the NLOS environment.}
environment are given as follows \cite{11}\cite{11r}.

\begin{equation}
\begin{aligned}
& P{L_{mmw}}(d)[dB] = -75.3 + 16.8\log (d) + {\cal N}(0,2.45),\\
& P{L_{2.4GHz}}(d)[dB] = -48.5 + 20.5\log (d) + {\cal N}(0,3.04).
\end{aligned}
\end{equation}

\begin{figure}[!t]
\centering
\includegraphics[width=0.48\textwidth]{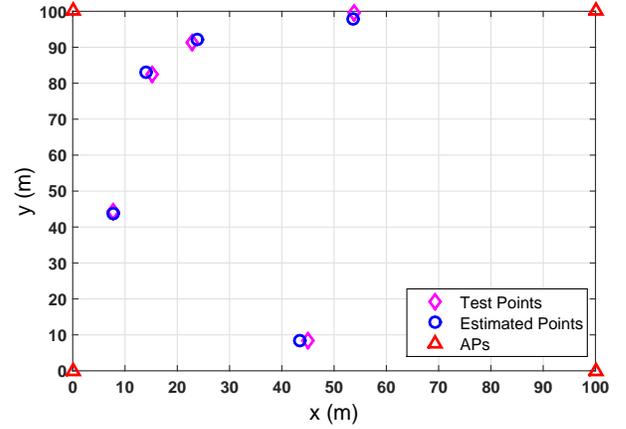}
\caption{A test of DoA-LF positioning algorithm.}
\label{f4}
\end{figure}

In order to create an
offline database in the training phase,
$100$ samples of RSSI and
DoA estimations at each RP are achieved.
Then the average RSSI and DoA estimations at each RP
are recorded in the offline database.
The size of indoor environment is
$100 \rm{m} \times 100 \rm{m}$
and RPs are deployed per $5$ meters.
Besides, $4$ APs are distributed
in the $4$ corners
of the entire region.
The transmit power of each AP
is $30$ mW.
Fig. \ref{f4} illustrates
the intuitive results
of DoA-LF algorithm, where the estimated
points match
the test points
very well.

In order to present the errors
of different positioning methods,
the probability of
successful positioning within the range of
location error, namely,
the cumulative distribution function (CDF)
of LF positioning is simulated.
Assume that the positioning error is $E$,
which is a random variable.
For any positioning
error $E = e$ in the horizontal axis,
its corresponding CDF,
namely, the value in vertical axis is the probability of $E < e$.
Besides,
the average positioning errors
of different methods are simulated.

\begin{figure}[!t]
\centering
\includegraphics[width=0.48\textwidth]{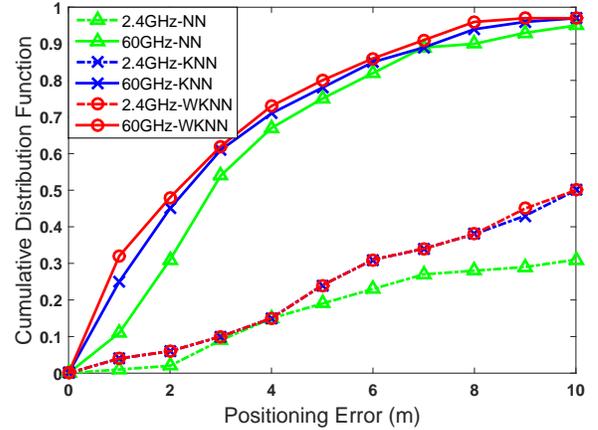}
\caption{Comparison of CDFs between 60 GHz mmWave and 2.4 GHz WiFi with different algorithms, where $K = 6$ for KNN and WKNN.}
\label{f5}
\end{figure}

\begin{figure}[!t]
\centering
\includegraphics[width=0.48\textwidth]{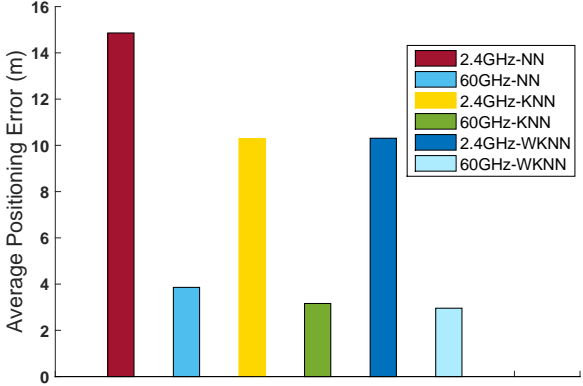}
\caption{Comparison of average positioning errors between 60 GHz mmWave and 2.4 GHz WiFi with different algorithms, where $K = 6$ for KNN and WKNN.}
\label{f6}
\end{figure}

Firstly, the impact of path loss models
of 2.4 GHz WiFi and 60 GHz mmWave
on three different LF positioning algorithms,
namely, the algorithms of NN, KNN and WKNN,
is illustrated in Fig. \ref{f5} and Fig. \ref{f6}.
In Fig. \ref{f5}, the
CDF curve of 60 GHz mmWave grows much faster
than that of 2.4 GHz WiFi when using
the same LF positioning
algorithm\footnote{The proposed DoA-LF algorithm also adopts WKNN operations. However,
in this section,
the algorithms of nearest node (NN),
$K$ nearest node (KNN) and weight $K$ nearest node
(WKNN) denote the LF positioning algorithms without
DoA estimation, namely, they only use RSSI information.}.
The underlying reason is that mmWave has
the characteristics of narrow beam and fast attenuation,
which can increase the discrimination of RSSI and
DoA information of different RPs and reduce the positioning error.
Besides, the positioning error
of 60 GHz mmWave
is much higher than that of 2.4 GHz WiFi
when using the same LF positioning algorithm.
Moreover, WKNN algorithm yields
the lowest error and
the fastest convergence
speed among the three LF
algorithms for 60 GHz mmWave.
Fig. \ref{f6} shows that
60 GHz mmWave yields
less positioning error compared
with 2.4 GHz WiFi.
The average positioning error
of 60 GHz mmWave is $68.9\%$ less than
2.4 GHz WiFi.

\begin{figure}[!t]
\centering
\includegraphics[width=0.48\textwidth]{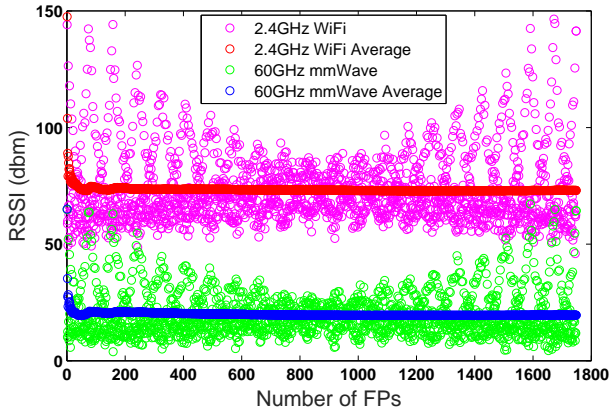}
\caption{Comparison of RSSI between 60 GHz mmWave and 2.4 GHz WiFi with the same transmit power of APs.}
\label{f7}
\end{figure}

Fig. \ref{f7} illustrates
the comparison of
RSSI
between 60 GHz mmWave and 2.4 GHz WiFi.
In Fig. \ref{f7}, the RSSI of
2.4 GHz WiFi is higher than that of
60 GHz mmWave at the same RP,
which verifies the fact
that the attenuation of 60 GHz mmWave signal
is faster than that of 2.4 GHz signal.
Besides, the distribution of
RSSIs of mmWave signal
is more concentrated compared
with 2.4 GHz signal, which can
increase the discrimination of RSSI
to reduce positioning error.

\begin{figure}[!t]
\centering
\includegraphics[width=0.48\textwidth]{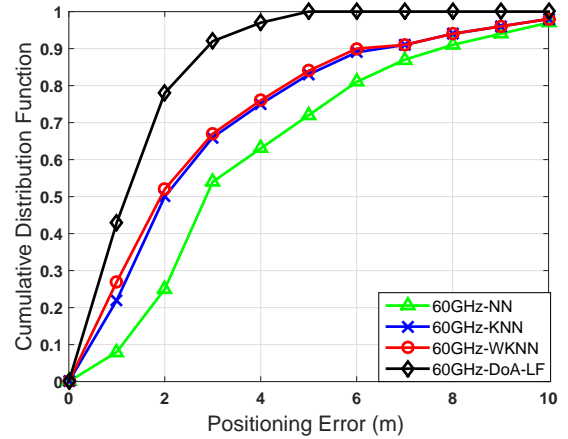}
\caption{Comparison of CDFs between DoA-LF algorithm and other algorithms in 60 GHz mmWave, where $K = 4$ for DoA-LF, KNN and WKNN.}
\label{f8}
\end{figure}

\begin{figure}[!t]
\centering
\includegraphics[width=0.48\textwidth]{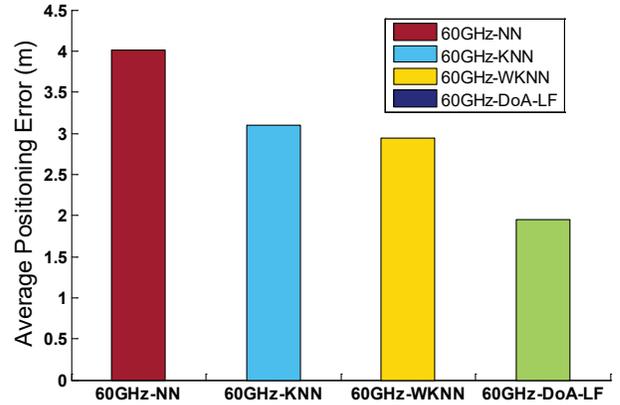}
\caption{Comparison of average positioning errors between DoA-LF algorithm and other algorithms in 60 GHz mmWave, where $K = 4$ for DoA-LF, KNN and WKNN.}
\label{f9}
\end{figure}

Then we analyze the performance of
the proposed DoA-LF algorithm which
uses RSSI and DoA information
for hybrid LF positioning with mmWave.
The performance is compared with other three LF
algorithms, which is illustrated
in Fig. \ref{f8} and Fig. \ref{f9}.
Fig. \ref{f8} shows that the
CDF curve of DoA-LF algorithm grows faster
than other algorithms without DoA estimation,
which means that the positioning error
of DoA-LF algorithm is the lowest.
Similarly, as illustrated in Fig. \ref{f9},
DoA-LF algorithm yields less
positioning error compared with other algorithms.
The average positioning error of
DoA-LF algorithm is 1.32 meters,
which is approximately $50\%$ less than WKNN
and KNN algorithm without DoA estimation.

\begin{figure}[!t]
\centering
\includegraphics[width=0.48\textwidth]{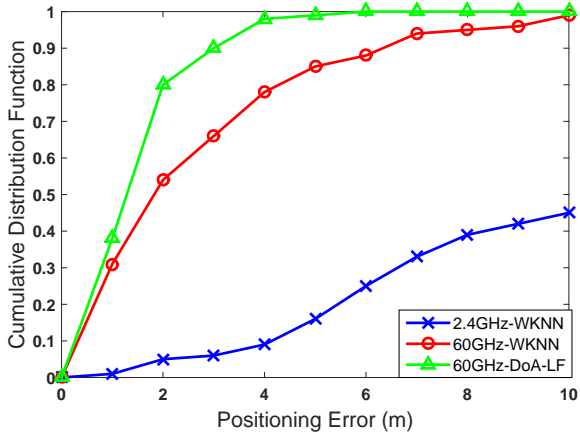}
\caption{Comparison of CDFs between DoA-LF with 60 GHz mmWave,
WKNN with 60 GHz mmWave and WKNN with 2.4GHz WiFi, where $K = 6$
for DoA-LF and WKNN.}
\label{f10}
\end{figure}

\begin{figure}[!t]
\centering
\includegraphics[width=0.48\textwidth]{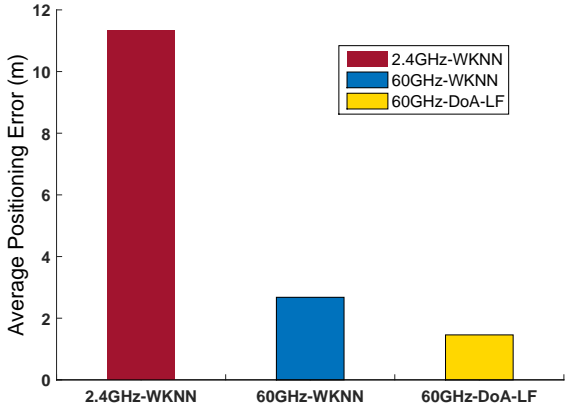}
\caption{Comparison of average positioning errors
between DoA-LF with 60 GHz mmWave, WKNN with 60 GHz mmWave
and WKNN with 2.4GHz WiFi, where $K = 6$
for DoA-LF and WKNN.}
\label{f11}
\end{figure}

Finally, we compare the
performance of DoA-LF algorithm with mmWave,
WKNN algorithm
with 60 GHz mmWave and WKNN algorithm with 2.4 GHz WiFi.
As illustrated in Fig. \ref{f10},
the positioning error of DoA-LF algorithm
with 60 GHz mmWave is much lower than WKNN algorithm
with 60 GHz mmWave and 2.4 GHz WiFi.
As illustrated in Fig. \ref{f11},
the average error distance of DoA-LF algorithm
is 1.37 meters, which is $48.5\%$
less than WKNN with 60 GHz mmWave and $85.6\%$
less than WKNN with 2.4 GHz WiFi.

\begin{figure}[!t]
\centering
\includegraphics[width=0.48\textwidth]{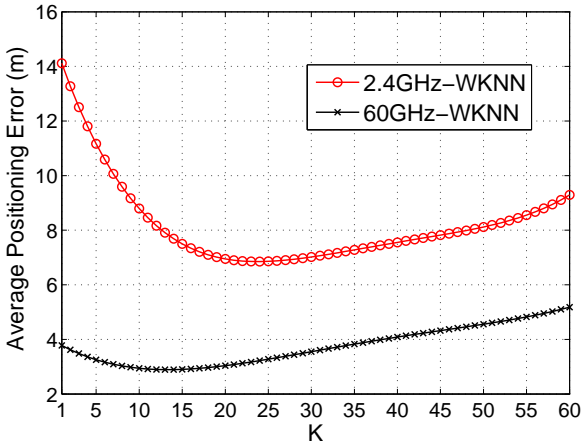}
\caption{The relation between the average positioning error and $K$.}
\label{fig_k}
\end{figure}

The relation between the average positioning error and $K$ is
illustrated in Fig. \ref{fig_k} to yield the
optimal value of $K$ in WKNN algorithm.
It is noted that if $K$ is small,
the randomness is large and the average positioning error is large.
On the contrary,
if $K$ is large, the
disturbance from other RPs is correspondingly large and the
average positioning error is also large.
Thus there exists an optimal $K$ to minimize the
average positioning error,
which is illustrated in Fig. \ref{fig_k}.
Besides, the
average positioning error for any $K$ with mmWave
is still smaller than that with low frequency signal.

Overall, we have verified that
the positioning error
can be reduced significantly
when employing DoA-LF positioning
algorithm with mmWave.

\subsection{CRLB and Analysis}

In this subsection,
the simulation results are shown to analyze
the impact of various parameters
on the average positioning
error of DoA-LF algorithm,
which include the number of APs,
the interval of RPs,
the path loss exponent,
the DoA estimation error and
the variance of received signal strength.
These simulations and analysis
can verify our analysis of CRLB
in section \ref{sec_CRLB}.
Besides, they can provide
a guideline for the selection of
appropriate parameters
in the proposed DoA-LF
positioning algorithm.

\begin{figure}[!t]
\centering
\includegraphics[width=0.48\textwidth]{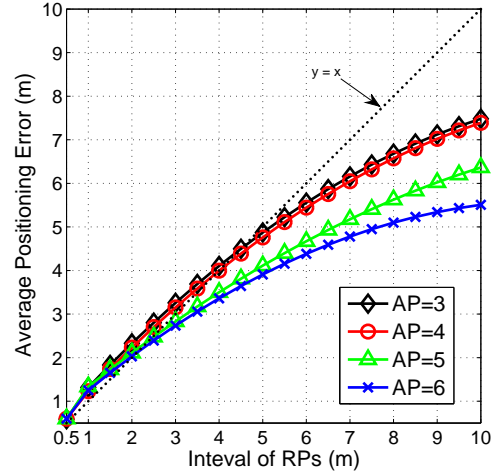}
\caption{The impact of the interval of RPs on average positioning error, where
 the number of APs varies from 3 to 6.}
\label{f12}
\end{figure}

Firstly, we analyze the
impact of the interval
of RPs on average positioning error.
Fig. \ref{f12} shows that
the average positioning error is increasing with the increase of the interval of RPs.
The curve $y = x$
is plotted in Fig. \ref{f12},
where $x$ denotes the interval of RPs and $y$ denotes
the average positioning error.
It is noted that when the interval of RPs is too small,
the average positioning error is larger than the
interval of RPs, which is due to the
fact that the CRLB has determined that the average positioning error cannot be
as small as possible.
Hence the interval of RPs does not have to be very small
because the computational complexity of DoA-LF
algorithm will be intolerable in this situation
and the average positioning error is lower bounded by the CRLB.
However,
when the interval of RP is larger than a threshold,
the average positioning error is smaller than the
interval of RPs.

\begin{figure}[!t]
\centering
\includegraphics[width=0.48\textwidth]{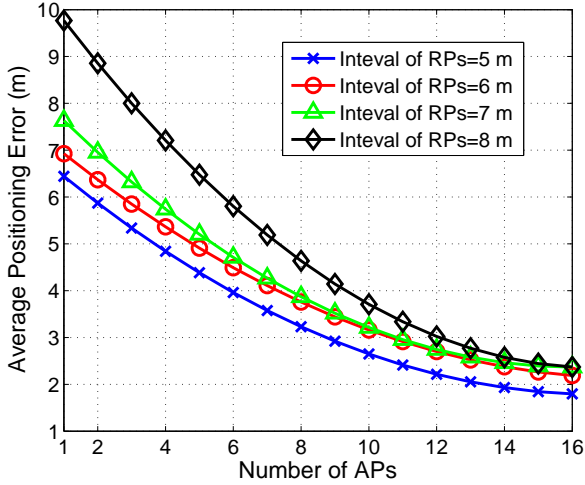}
\caption{The impact of the number of APs on average positioning error,
where the interval of RPs varies from 5 to 8 meters.}
\label{f13}
\end{figure}

Then we analyze the impact
of the number of APs
on average positioning error.
As illustrated in Fig. \ref{f13},
the positioning error is decreasing
with the increase of the number of APs.
Since the increase of the number of APs will enlarge the
computational complexity,
there is an optimal number of APs in order to
balance the computational complexity and
positioning error.

\begin{figure}[!t]
\centering
\includegraphics[width=0.48\textwidth]{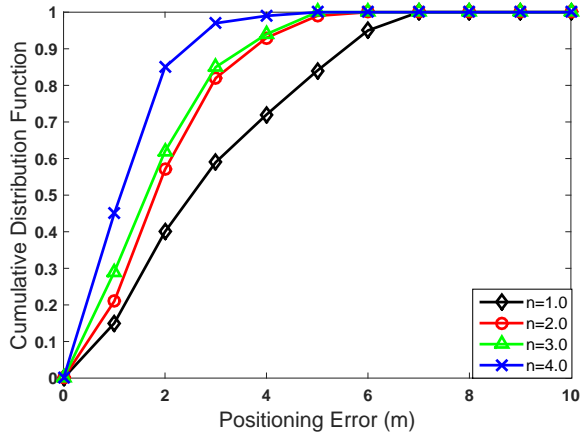}
\caption{The relation between positioning error and path loss exponent.}
\label{f14}
\end{figure}

\begin{figure}[!t]
\centering
\includegraphics[width=0.48\textwidth]{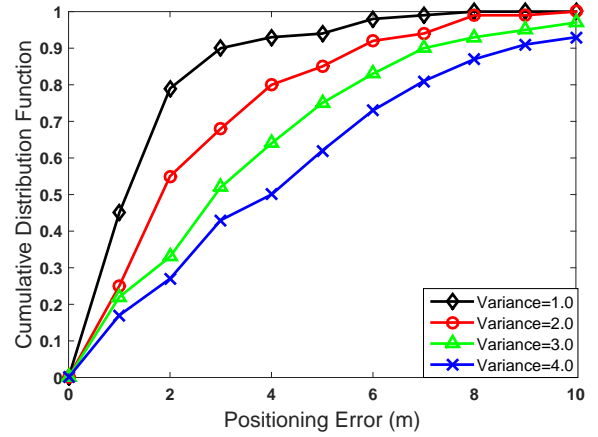}
\caption{The relation between positioning error and received signal strength variance.}
\label{f15}
\end{figure}

\begin{figure}[!t]
\centering
\includegraphics[width=0.48\textwidth]{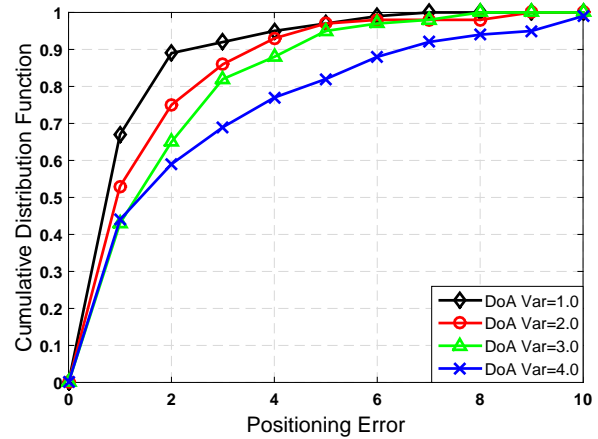}
\caption{The relation between positioning error and DoA estimation variance.}
\label{f16}
\end{figure}

Finally, we analyze the impact of
the variance of DoA estimation,
the path loss exponent and the
variance of received signal strength
on the average positioning error
of DoA-LF algorithm with mmWave.
In Fig. \ref{f14}, the positioning error is
decreasing with the
increase of path loss exponent.
In Fig. \ref{f15}, the positioning error is
decreasing with the decrease
of the variance of
received signal strength.
In Fig. \ref{f16}, the positioning error is decreasing with the decrease of
the variance of DoA estimation.
These results support our analysis
results of CRLB in Section \ref{sec_CRLB},
which can be explained.
When the path loss exponent is large,
the discrimination of RSSI
is correspondingly large, which can
reduce positioning error.
Besides, when the variance of signal intensity
or DoA estimation is small,
the fluctuation of RSSI or
DoA estimation is correspondingly small,
which can also reduce positioning error.
However, the path loss exponent,
the variance of received signal strength and
the variance of DoA estimation are
not controllable variables because they are
impacted by the electromagnetic
and physical environment.
Therefore when the environment is not ideal,
we can properly reduce the interval of RPs or
increase the number of APs to
bound positioning error.

\section{Conclusion}

In this paper, the DoA-LF positioning algorithm
is proposed, which incorporates the features
of RSSI and DoA information into LF positioning.
The characteristics of narrow beam and
fast attenuation of mmWave are
exploited
to reduce positioning error.
The DoA information obtained from MUSIC algorithm
is combined with RSSI
information to construct a joint
offline database
for online matching.
Then the $K$ candidate RPs with the most similar features
are selected
and
their weighted mean is calculated,
which is the estimated location.
Moreover, the CRLB of DoA-LF
positioning algorithm with
mmWave is derived.
Simulation results show that the positioning error of
DoA-LF algorithm
is much smaller than the
algorithms without DoA estimation.
Besides,
the positioning error
with mmWave is much
smaller than that with low frequency signals.
Moreover, we have shown that there exists
an optimal $K$ to minimize the positioning error via simulation.
Finally, simulation results show that the positioning error
is an increasing function of the shadow
fading factor, the interval of RPs and the
error of DoA estimation.
And the positioning error
is a decreasing function of
the path loss exponent and the number of APs.
The study of this paper may provide
guideline for indoor positioning
in the era of
5G with densely deployed
mmWave small cells.

\section*{Acknowledgment}
The authors appreciate editor and anonymous reviewers for
their precious time and great effort in improving this paper.

\end{document}